\documentclass[aps,preprint,tightenlines]{revtex4}
\usepackage{graphicx} 

\begin{document}

\title{Hard breakup of the deuteron into two $\Delta$ -isobars}

\author{Carlos~G. Granados and Misak~M.~Sargsian}

\affiliation{Florida International University, Miami, FL 33199 USA}

\date{\today}

\begin{abstract}
We study high energy photodisintegration of the deuteron into two  $\Delta$-isobars at large 
center of mass angles within the QCD hard rescattering 
model (HRM). According to the HRM, the process develops  in three main steps: the photon knocks a quark from one of 
the nucleons in the deuteron; the struck quark rescatters off a quark from  the other nucleon sharing the high  energy 
of the photon; then the energetic quarks recombine into two outgoing baryons which have large transverse momenta.
Within the HRM, the cross section is expressed through the amplitude of $pn\rightarrow \Delta\Delta$ scattering which 
we evaluated based on the quark-interchange model of hard hadronic scattering.   Calculations show that 
the angular distribution and the strength of the photodisintegration  is mainly determined by the properties of the 
 $pn\rightarrow \Delta\Delta$  scattering.   We predict that the cross section of the  deuteron 
breakup to $ \Delta^{++}\Delta^{-}$ is 4-5 times larger  than that of the breakup  to  the $ \Delta^{+}\Delta^{0}$ channel.  Also, the angular distributions for 
these two channels are markedly different.   These  can be compared with the predictions  based on 
the assumption that two hard $\Delta$-isobars are the  result of the disintegration of the preexisting $\Delta\Delta$ components 
of the deuteron wave function.  In this case, one expects  the  angular distributions and cross sections of the 
breakup in both $ \Delta^{++}\Delta^{-}$  and  $ \Delta^{+}\Delta^{0}$ channels to be similar.
 \end{abstract}
\maketitle

\section{Introduction}
Hard nuclear processes  provide an important testing ground for QCD degrees of freedom in nuclei. 
One of such processes is the high-energy large-angle photodisintegration of light nuclei.
These reactions were intensively studied during the last two decades. The studies included the experiments 
on large center of mass~(c.m.) angle  break-up of the deuteron into the $pn$ 
pair\cite{NE8,NE17,E89012,Schulte1,gdpnpolexp1,Schulte2,Mirazita,gdpnpolexp2} 
as well as break-up of  the $^3$He nucleus  into two high energy protons and a slow  neutron\cite{Pomerantz:2009sb}. 
The uniqueness of these reactions is in the  effectiveness by which high momentum and energy are transferred to 
the NN system\cite{Gilman:2001yh} at a given photon energy, $E_\gamma$. Namely at  large and fixed values of the c.m. scattering 
angle,   $s,-t\sim 2M_dE_\gamma$ which is by a factor of two larger than the invariant energy 
and transfered momenta achieved  in hadronic interactions at the same incident energies.

The above mentioned reactions confirmed the prediction of quark-counting rule\cite{BCh} according to which the 
energy  dependence of the differential cross section at  large c.m. scattering angles  scales 
as ${d\sigma\over dt} \sim s^{-11}$.

However, calculations of the absolute cross sections  require a more detailed 
understanding of the dynamics of these processes.  The considered theoretical models can  be 
grouped  by two distinctly different underlying assumptions made in the calculations\cite{Brodskyetal}. 
The first assumes that the large c.m. angle  nucleons are produced through the interaction of  the incoming  
photon with a pre-existing hard two nucleon system in the nucleus\cite{RNA1,RNA2,DN}.  
The second approach is based on the assumption that the two high  momentum  nucleons are produced  
through a hard rescattering at the final state of the 
reaction\cite{QGS,Frankfurt:1999ik,Frankfurt:1999ic,Sargsian:2008ez,Sargsian:2003sz,Sargsian:2008zm}.

In the hard rescattering model (HRM)\cite{Frankfurt:1999ik} in particular, by explicitly introducing 
quark degrees of freedom, a parameter free cross section has been obtained for hard 
photodisintegration of the  deuteron at 90$^0$ c.m.   angle \cite{Frankfurt:1999ik,Frankfurt:1999ic}.  Also HRM's  prediction of the  hard breakup of 
two protons from the $^3$He nucleus\cite{Sargsian:2008zm} agreed reasonably well with the recent experimental 
data\cite{Pomerantz:2009sb}.

In the present work we extend the  HRM approach to calculate hard breakup of the deuteron into   
two-$\Delta$-isobars  produced at large angles in the $\gamma-d$ center of mass reference frame.
In our estimates, we calculate the relative strength of $\gamma d\rightarrow \Delta^{++}\Delta^{-}$ and 
 $\gamma d\rightarrow \Delta^{+}\Delta^{0}$ cross sections as they compare with the   $\gamma d\rightarrow pn$ cross 
section.

The investigation of the production of two energetic $\Delta$-isobars from the deuteron has an important significance in probing 
possible non-nucleonic components in the deuteron wave function. 
Studies of non-nucleonic components of the deuteron have a rather long history.  Already in the 1970's,  the possible 
existence of the baryonic resonance components in the deuteron have been studied in potential and pion-exchange models 
(see e.g.\cite{Nath:1971ts,Arenhovel:1971de,Benz:1974au,Benz:1974au,Rost:1975zn}). They were also considered 
in quark-interchange\cite{Glozman:1994xe} and chiral quark\cite{JuliaDiaz:2002gu} models.

Among the all possible resonance components  the $\Delta\Delta$ component has an  
interesting relation to the possible existence of  the hidden color component in the deuteron wave function
( see e.g. Refs.\cite{Harvey:1980rva,Brodsky:1983vf,Brodsky:1985gs,FS88,SRCRev}). 
This relation follows from the observation\cite{Harvey:1980rva,Brodsky:1983vf} that  in the 
regime in which the chiral symmetry is restored  the color singled 6-quark configuration can be 
expressed through the superposition of $NN$, $\Delta\Delta$ and hidden-color components with
relative normalizations  fixed by the $SU(3)$ symmetry.   Thus, experimental verification of the relative 
strength of the $NN$ to $\Delta\Delta$ component could shed light on the existence of hidden color 
components in the deuteron wave function. However, both components should be probed in 
hard nuclear  processes in which case small inter-nucleon distances in the deuteron are  probed.
Our calculation in this case will allow us to asses the role of the hard rescattering in these processes. 
It will allow us also to explore another venue for checking the basic mechanism of the high momentum transfer  
breakup of nuclei into two baryons. Our calculations result in the distinct predictions for angular 
distributions of the $\Delta$-isobar pair at large c.m. production angle as well as their relative strength compared with  
the production of the $pn$ pair at the same kinematics.  
Despite  experimental challenges associated with the investigation of 
two $\Delta$-isobar  breakup of the deuteron\cite{PR}, there are ongoing efforts in performing 
such experiment at  Jefferson Lab\cite{Gao,DM} which we hope will allow to verify our predictions.

\section{Hard Rescattering Model}
\label{hrm}
We consider the  photoproduction of two baryons, $B_1$ and $B_2$, in the reaction,
\begin{equation}
\gamma + d \rightarrow B_1 + B_2
\label{Reaction}
\end{equation}
in which the  baryons are produced at large angles in the $\gamma-d$ center of mass reference frame.

According to  the HRM,  the  large angle breakup of the  NN system proceeds through the knock-out of 
a valence quark from one of the nucleons with subsequent hard rescattering of the  struck-quark
with a valence quark of the  second nucleon.  
 The two quarks then recombine with the spectator systems of nucleons forming  
two emerging baryons with large transverse momenta. The hard rescattering provides the mechanism 
of sharing  the photon's energy among two final baryons.
 
\begin{figure}[ht]
\centering\includegraphics[height=4cm,width=6cm]{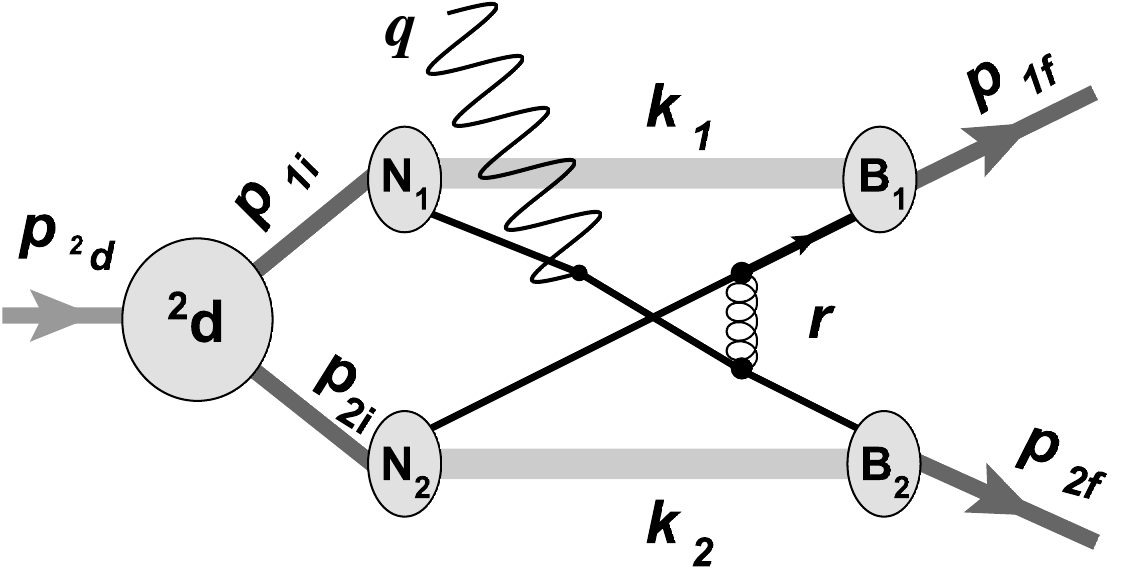}
\caption{Deuteron photodisintegration according to the HRM}
\label{hrmf}
\end{figure}

The invariant amplitude of  the photodisintegration Eq.(\ref{Reaction})  is calculated 
by applying Feynman diagram  rules to diagrams similar to  Fig.\ref{hrmf}.  During the calculation we introduce 
 undetermined quark wave functions of baryons to account for the transition of the initial nucleons to 
 the quark-spectator systems, and also for the recombination of the final state quarks with these spectator systems into 
the final two baryon system.

Fig.\ref{hrmf} displays the chosen independent momenta for three loop integration involved in the 
invariant amplitude.  Two major approximations  simplify further calculations. First, using the fact that 
the struck quark is very energetic we treat it on its mass shell. Then the struck quark's propagator is evaluated
at it's pole value at such magnitudes of  nucleon momenta that maximize the deuteron wave function. 
These approximations allow us to factorize the invariant amplitude into three distinguished parts.  The first, 
representing the transition amplitude of the deuteron into the ($pn$) system, which can be evaluated using a
realistic deuteron wave function.  The second is the amplitude of photon-quark interaction, and  the third term 
represents the hard rescattering of the struck quark with  recombination into a  two large transverse momentum baryonic system.
Combined with the initial state nucleon wave functions, the rescattering part is expressed through the 
quark-interchange~(QI) amplitude of $pn\rightarrow B_1 B_2$ scattering. Details of the derivation are given 
in  Refs.\cite{Frankfurt:1999ik,Sargsian:2008zm}.
After the  above mentioned factorization is made, the overall invariant amplitude of $\gamma d\rightarrow B_1 B_2$ 
reaction can be expressed as follows:
\begin{eqnarray}
& & \langle\lambda_{1f},\lambda_{2f}\mid {\cal M} \mid \lambda_\gamma, \lambda_d\rangle 
 =  ie[\lambda_\gamma]\times  \nonumber \\
& & \ \ \ \ \left\{ \sum\limits_{i \in N_1}\sum\limits_{\lambda_{2i}} \int 
{Q_i^{N_1}\over \sqrt{2s'}}
\langle \lambda_{2f};\lambda_{1f}\mid T^{QI}_{(pn\to B_1B_2),i}(s,t_N)\mid 
\lambda_\gamma;\lambda_{2i}\rangle 
\Psi_{\textnormal{\scriptsize {d}}}^{\lambda_d}(p_{1i},\lambda_\gamma;p_{2i},\lambda_{2i})
{d^2p_\perp \over (2\pi)^2} \right. \nonumber \\ 
&  &\ \ \ \  + \left.
\sum\limits_{i \in N_2}\sum\limits_{\lambda_{1i}} \int 
{Q_i^{N_2}\over \sqrt{2s'}}
\langle \lambda_{2f};\lambda_{1f}\mid T^{QI}_{(pn\to B_1B_2),i}(s, t_{N})\mid \lambda_{1i};
\lambda_{\gamma}\rangle 
\Psi_{\textnormal{\scriptsize {d}}}^{\lambda_d}(p_{1i},\lambda_{1i};p_{2i},\lambda_{\gamma})
{d^2p_\perp \over (2\pi)^2} \right\}
\label{ampl}
\end{eqnarray}
where $\lambda_\gamma$,$\lambda_d$, $\lambda_{1f}$ and $\lambda_{2f}$ are the helicities of the
photon, deuteron and the two outgoing baryons respectively.
Here $\Psi_{\textnormal{\scriptsize {d}}}^{\lambda_d}(p_{1i},\lambda_{1i};p_{2i},\lambda_{2i})$ is the $\lambda_d$-helicity 
light-cone deuteron wave function defined in the $q_+=0$ reference frame.  The initial light-cone momenta of 
the nucleons in the deuteron are $p_{1i}=(\alpha_{1i}={1\over 2}, p_{1i\perp}=-p_\perp)$ and 
$p_{2i} = (\alpha_{2i}={1\over 2},p_{2i\perp}=p_\perp)$ with $\lambda_{1i}$ 
and $\lambda_{2i}$ being their helicities respectively.  
The ${1\over \sqrt{s^\prime}}$ factor with $s^\prime = s-M_d^2$ comes from the energetic propagator 
of the struck quark before its rescattering.  The squares of the total invariant energy as well as the momentum 
transfer are defined as follows:
\begin{eqnarray}
s & = & (q + p_d)^2 = (p_{1f} + p_{2f})^2 = 2E^{lab}_\gamma M_d + M_d^2 \nonumber \\
t &  = & (p_{1f}-q)^2 = (p_{2f}-p_d)^2 
\label{kin}
\end{eqnarray}
where $q$, $p_d$, $p_{1f}$ and $p_{2f}$ are the four-momenta of the photon, deuteron and two outgoing baryons respectively. 
The lab energy of the photon is defined by $E^{lab}_\gamma$, and $M_d$ is  the mass of the deuteron.
The transfer momentum, $t_N$ in the rescattering amplitude in Eq.(\ref{ampl}) is defined as:
\begin{equation}
t_N = (p_{1f}-p_{1i}-q)^2 = (p_{2f}-p_{2i}) \approx   (p_{2f}-{p_d\over 2})^2  = {t\over 2} + 
{m_{B2}^2\over 2} - {M_d^2\over 4},
\label{tN}
\end{equation}
where the approximation in the right hand side follows from the assumption that the magnitudes of light-cone momentum fractions 
of  bound nucleons dominating in the scattering amplitude are $\alpha_{1i}=\alpha_{2i} = {1\over 2}$, and that the transverse 
momenta of these nucleons are negligible as compared to the momentum transfer in the reaction, $p_\perp^2 \ll |t_N|,|u_N|$.

In Eq.(\ref{ampl})  the following expression
\begin{equation}
Q_i\langle \lambda_{2f};\lambda_{1f}\mid T^{QI}_{(pn\to B_1B2),i}(s,t_N)\mid 
\lambda_{1i};\lambda_{2i}\rangle 
\label{QT}
\end{equation}
represents the quark-charge weighted QI amplitude of $pn\rightarrow B_1 B_2$ hard exclusive 
scattering. The  factor   $Q_i$ corresponds to the charge (in $e$ units) of the quark  that interacts with the incoming photon. 
In a further approximation we factorize the hard rescattering amplitude  from the integral since the momentum transfer 
entering in $T_{(pn\to B_1B_2),i}(s, t_N)$ significantly exceeds the Fermi momentum of the nucleon in the deuteron. 
Also, after calculating the overall quark-charge factors, the QI scattering amplitudes are identified with the $NN\rightarrow B_1B_2$ helicity amplitudes as follows:
\begin{equation}
\langle \lambda_{2f};\lambda_{1f}\mid T_{pn\to B_1B_2}^{QI}(s,t_N)\mid \lambda_{1i};\lambda_{2i}\rangle =\phi_j(s,\theta_{c.m.}^{N}),
\label{notn}
\end{equation}
where $\theta_{c.m.}^N$ is the effective center of mass angle defined for given  $s$ and $t_N$.

\medskip

The differential cross section for unpolarized scattering is obtained through:
\begin{eqnarray}
{d\sigma_{\gamma d\rightarrow B_1B_2}\over dt}={1\over16\pi}{1\over (s-M_d^2)}|\bar{\cal M}|_{\gamma d\rightarrow 
B_1B_2}^2
\label{gdpnBB}
\end{eqnarray}
where
\begin{equation}
|\bar{\cal M}|_{\gamma d\rightarrow B_1B_2}^2 = 
{1\over 3}{1\over 2} \sum\limits_{\lambda_{1f},\lambda_{2f},\lambda_\gamma,\lambda_d}
\mid \langle \lambda_{1f},\lambda_{2f}\mid {\cal M}\mid \lambda_\gamma,\lambda_d\rangle|^2,
\label{M2}
\end{equation}
with  the invariant amplitude square  averaged by the number  of helicity states of the deuteron and photon.

\section{Cross section of the $\gamma + d\rightarrow pn$ breakup reaction}
\label{SECgdpn}

We derive the amplitude of the breakup of the deuteron into the $pn$ pair from Eq.(\ref{ampl}) by introducing the
independent helicity amplitudes of $pn$ elastic scattering Eq.(\ref{pnham}) and by separating the quark-charge factors into $\hat Q^{N_1}$ and $\hat Q^{N_2}$ 
which correspond to the scattering of the  photon off  the quark of the first and the second nucleons in the deuteron. Then, for Eq.(\ref{M2}) one obtains:
\begin{eqnarray}
\bar{|{\cal M}|^2}&=&{1\over2}{1\over3}{e^2\over 2s^\prime}\left[S_{12}\left\{|(\hat{Q}^{N_1}+\hat{Q}^{N_2})\phi_1|^2+
|(\hat{Q}^{N_1}+\hat{Q}^{N_2})\phi_2|^2\right\}\right.\nonumber\\
&&+\left.S_{34}\left\{|\hat{Q}^{N_1}\phi_3+\hat{Q}^{N_2}\phi_4|^2+|\hat{Q}^{N_1}\phi_4+
\hat{Q}^{N_2}\phi_3|^2\right\}\right.\nonumber\\
&&+\left.2S_0|(\hat{Q}^{N_1}+\hat{Q}^{N_2})\phi_5|^2\right],
\label{asamp}
\end{eqnarray}
where the light-cone spectral functions of the deuteron  are defined as follows:
\begin{eqnarray}
S_{12}&=&\sum^{1}_{\lambda=-1}\sum^{1\over2}_{(\lambda_1=\lambda_2=-{1\over2})}
\left|\int\Psi_{\textnormal{\scriptsize {d}}}^{\lambda_d}(p_1,\lambda_1;p_2,\lambda_2)
{d^2p_\perp \over (2\pi)^2}\right|^2,\nonumber\\
S_{34}&=&\sum^{1}_{\lambda=-1}\sum^{1\over2}_{(\lambda_1=-\lambda_2=-{1\over2})}\left
|\int\Psi_{\textnormal{\scriptsize {d}}}^{\lambda_d}(p_1,\lambda_1;p_2,\lambda_2)
{d^2p_\perp \over (2\pi)^2}\right|^2,\nonumber\\
S_0&=&S_{12}+S_{34}.
\label{spfun}
\end{eqnarray}
Eq.(\ref{asamp}) can be further simplified if we assume (see e.g.\cite{RS})  that $\phi_3\approx \phi_4$, 
as well as  $S_{12} \approx S_{34}  = {S_0\over 2}$, which results in:

\begin{eqnarray}
\bar{|{\cal M}|^2}&=&{1\over2}{1\over3}{e^2\over 2s^\prime}Q_{F,pn}^2 {S_0\over 2} \left[ 
|\phi_1|^2+|\phi_2|^2+ |\phi_3|^2 +|\phi_4|^2+|\phi_4|^2+
4|\phi_5|^2\right].
\label{asamp_2}
\end{eqnarray}
Using the  expression of the differential cross section of elastic $pn$ scattering:
\begin{equation}
{d\sigma^{NN\rightarrow NN}(s,\theta_{c.m.}^N)\over dt} =  {1\over 16\pi}{1\over s(s-4m_N^2)}
{1\over 2}(|\phi_1|^2 + |\phi_2|^2 + |\phi_3|^2 + |\phi_4|^2 + 4|\phi_5|^2),
\label{crs_NN}
\end{equation}
and the relation between  the light-cone   and   non-relativistic deuteron wave functions\cite{FS81,Frankfurt:1999ik,taggs,polext} at small internal momenta:
$\Psi_d(\alpha, p_\perp) = (2\pi)^{3\over 2}\Psi_{d,NR}(p)\sqrt{m_N}$  in Eq.(\ref{asamp}), 
for the  differential  cross section on obtains from Eq.(\ref{gdpnBB}):
\begin{equation}
{d\sigma^{\gamma d \rightarrow pn}(s,\theta_{c.m.}) \over dt} = {\alpha Q_{F,pn}^2 8\pi^4\over s^\prime} 
{d\sigma^{pn\rightarrow pn}(s,\theta_{c.m.}^N)\over dt} \bar S_{0,NR},
\label{gdpncrs}
\end{equation}
where  we neglected the difference between $4m_N^2$ and $M_d^2$. 
Here the  averaged non relativistic spectral function of  the deuteron is defined as follows:
\begin{equation}
\bar S_{0,NR} = {1\over 3}\sum\limits_{\lambda=-1}^{\lambda=1}
\sum\limits_{\lambda_1,\lambda_2=-{1\over2}}^{{1\over 2}}
\left|\int\Psi_{\textnormal{\scriptsize {d,NR}}}^{\lambda_d}(\alpha={1\over 2},p_\perp,\lambda_1;
\alpha={1\over 2},-p_{\perp},\lambda_2)\sqrt{m_N}
{d^2p_\perp\over (2\pi)^2}\right|^2,
\label{avspfun}
\end{equation}
where $\Psi_{d,NR}$ is the non relativistic deuteron wave function, which can be 
calculated using realistic $NN$ interaction potentials.  

The quark-charge factor, $Q_{F,pn} ={1\over 3}$\cite{Frankfurt:1999ik} 
accounts for the amount of the effective charge exchanged between the proton and the neutron in the 
rescattering.  It is estimated by counting all  the possible quark-exchanges within  the $pn$ pair weighted with 
the charge of one of the exchanged quarks  (for more details see Appendix B). The result in Eq.(\ref{gdpncrs}) is remarkably
simple and contains no free parameters. It can be evaluated using the experimental values of the differential 
cross section of the elastic $pn$ scattering, ${d\sigma^{pn\rightarrow pn}(s,\theta_{c.m.}^N)\over dt}$. 
The angle $\theta^{N}_{c.m.}$  entering in the $pn\rightarrow pn$ cross section is the center of mass angle of the scattering corresponding to the $NN$ 
elastic reaction at $s$ and $t_{N}$.  It is related to $\theta_{c.m.}$ of the $pn$ photodisintegration by \cite{Sargsian:2008zm}:
\begin{equation}
cos(\theta_{c.m.}^N) = 1 - {(s-M_d^2)\over 2(s-4m_N^2)}
{(\sqrt{s}-\sqrt{s-4 m_N^2}cos(\theta_{c.m.}))\over \sqrt{s}} + {4m_N^2 - M_d^2\over 2(s-4m_N^2)} .
\label{theta_cmn}
\end{equation} 
It is worth mentioning that as it follows from the  equation above, $\theta_{c.m.} = 90^0$ photodisintegration
will correspond to the $\theta_{c.m.}^{N} = 60^0$ hard $pn$ elastic rescattering at the final state of the reaction.

\section{Cross section of the $\gamma  d\rightarrow \Delta\Delta$ breakup reaction}
\label{delpr}

We use an approach similar to that in Sec.\ref{SECgdpn}   to derive the invariant  amplitude  of the 
 $\gamma  d\rightarrow \Delta\Delta$  reactions.  In this case Eq.(\ref{ampl}) requires an input of the helicity amplitudes 
of the corresponding  $pn\rightarrow\Delta\Delta$ scattering.  One has a total 32 independent 
helicity amplitudes for this scattering.  To simplify further our  derivations, we will restrict ourselves 
by considering only  the seven helicity conserving amplitudes given in Eq.(\ref{hconamp}).  Using these 
amplitudes in Eq.(\ref{ampl}) and separating the quark-charge factors into $\hat Q^{N_1}$ and $\hat Q^{N_2}$,  similar to Eq.(\ref{asamp}) one obtains
 \begin{eqnarray}
\bar{|{\cal M}|^2}_{\gamma d\rightarrow\Delta\Delta}&=&{1\over2}{1\over3}{e^2\over 2s^\prime}\left[S_{12}\left\{|(\hat{Q}^{N_1}+\hat{Q}^{N_2})\phi_1|^2
+|(\hat{Q}^{N_1}+\hat{Q}^{N_2})\phi_6|^2+|(\hat{Q}^{N_1}+\hat{Q}^{N_2})\phi_7|^2\right\}\right.\nonumber\\
&&+\left.S_{34}\left\{|\hat{Q}^{N_1}\phi_3+\hat{Q}^{N_2}\phi_4|^2+|\hat{Q}^{N_1}\phi_4+\hat{Q}^{N_2}\phi_3|^2\right.\right.\nonumber\\
&&+\left.\left.|\hat{Q}^{N_1}\phi_8+\hat{Q}^{N_2}\phi_9|^2+|\hat{Q}^{N_1}\phi_9+\hat{Q}^{N_2}\phi_8|^2\right\}\right],
\label{asampDD}
\end{eqnarray}
where $S_{12}$ and $S_{34}$ are defined in Eq.(\ref{spfun}). Similar to the previous section, 
we simplify further the above expression assuming that all helicity conserving amplitudes are of the 
same order of magnitude. Assuming also that  $S_{12}\approx S_{34} \approx {S_0\over 2}$, we obtain
\begin{eqnarray}
\bar{|{\cal M}|^2}&=&{1\over2}{1\over3}{e^2\over 2s^\prime} Q_{F,\Delta\Delta} {S_0\over 2} \left[ 
|\phi_1|^2+|\phi_3|^2 +|\phi_4|^2+|\phi_4|^2
+|\phi_6|^2+|\phi_7|^2+|\phi_8|^2+
|\phi_9|^2\right],
\label{asampDD_2}
\end{eqnarray}
where $Q_{F,\Delta\Delta} = \hat Q^{N_1} +  \hat Q^{N_2} = {1\over 3}$  is obtained by using the same approach as for the case of 
the $pn$ breakup in Sec.\ref{SECgdpn}.   Using now the expression of the 
differential cross section of  $pn\rightarrow\Delta\Delta$ scattering,
\begin{eqnarray}
& & {d\sigma^{pn\rightarrow\Delta\Delta}({s,\theta^N_{c.m.}})\over dt}= \nonumber \\
& & \ \ \ \ {1\over16\pi}{1\over(s-4m^2_N)}{1\over2} \left[ |\phi_1|^2+|\phi_3|^2 +|\phi_4|^2+|\phi_4|^2
+|\phi_6|^2+|\phi_7|^2+|\phi_8|^2+ |\phi_9|^2\right] 
\label{crs_pn_DD}
\end{eqnarray}
as well as the relation between light-cone and non relativistic deuteron wave function  discussed in Sec.\ref{SECgdpn}, from 
Eq.(\ref{gdpnBB}) we obtain the following expression for the differential cross section of the
$\gamma d\rightarrow \Delta\Delta$ scattering:
 \begin{equation}
{d\sigma^{\gamma d \rightarrow \Delta\Delta}(s,\theta_{c.m.}) \over dt} = {\alpha Q_{F,\Delta\Delta}^2 8\pi^4\over s^\prime} 
{d\sigma^{pn\rightarrow \Delta\Delta}(s,\theta_{c.m.}^N)\over dt} \bar S_{0,NR},
\label{gdDDcrs}
\end{equation}
where $\bar S_{0,NR}$ is given in  Eq.(\ref{avspfun}).  The  effective c.m. angle $\theta_{c.m.}^N$ entering in 
the argument of the differential cross section of $pn\rightarrow \Delta\Delta$ reaction can  be calculated
by using Eqs. (\ref{kin}) and (\ref{tN}) to obtain
\begin{equation}
cos\theta^N_{c.m.}={1\over\sqrt{\left(s-4m_N^2\right)\left(s-4m_\Delta^2\right)}}\left[s-{M_d^2-4m_N^2\over2}-{s-M_d^2\over2
\sqrt s}\left(\sqrt s-\sqrt{s-4m_\Delta^2}cos\theta_{c.m.}\right)\right].
\label{kin2}
\end{equation}
 
As it follows from Eq.(\ref{gdDDcrs}),  provided there are  enough experimental data on high momentum transfer 
 $pn\rightarrow\Delta\Delta$ differential  cross sections, the ${\gamma d \rightarrow \Delta\Delta}$ cross section 
can be computed without introducing  an adjustable free parameter.  However,  there are no experimental data on  hard exclusive
$pn\rightarrow \Delta\Delta$ reactions with sufficient accuracy that would allow us  to make   quantitative 
estimates  based on Eq.(\ref{gdDDcrs}).  Instead, in the next section we will attempt  to make quantitative 
predictions  based on the  quark-interchange framework of   hard scattering.

\section{Estimates of the relative strength of the  $\Delta\Delta$ breakup reactions.}

Our further calculations are  based on the experimental observation\cite{h20} that 
the  quark-interchange\cite{Sivers:1975dg}  represents the dominant mechanism of  hard exclusive scattering of baryons 
that carry valence quarks with common flavor.   The  quark-interchange mechanism however will not allow
us to calculate the absolute cross sections.  Instead, we expect that its predictions 
will be more reliable for the ratios of the differential cross sections for different exclusive channels.

\begin{figure}[ht]
\centering\includegraphics[height=6cm,width=8cm]{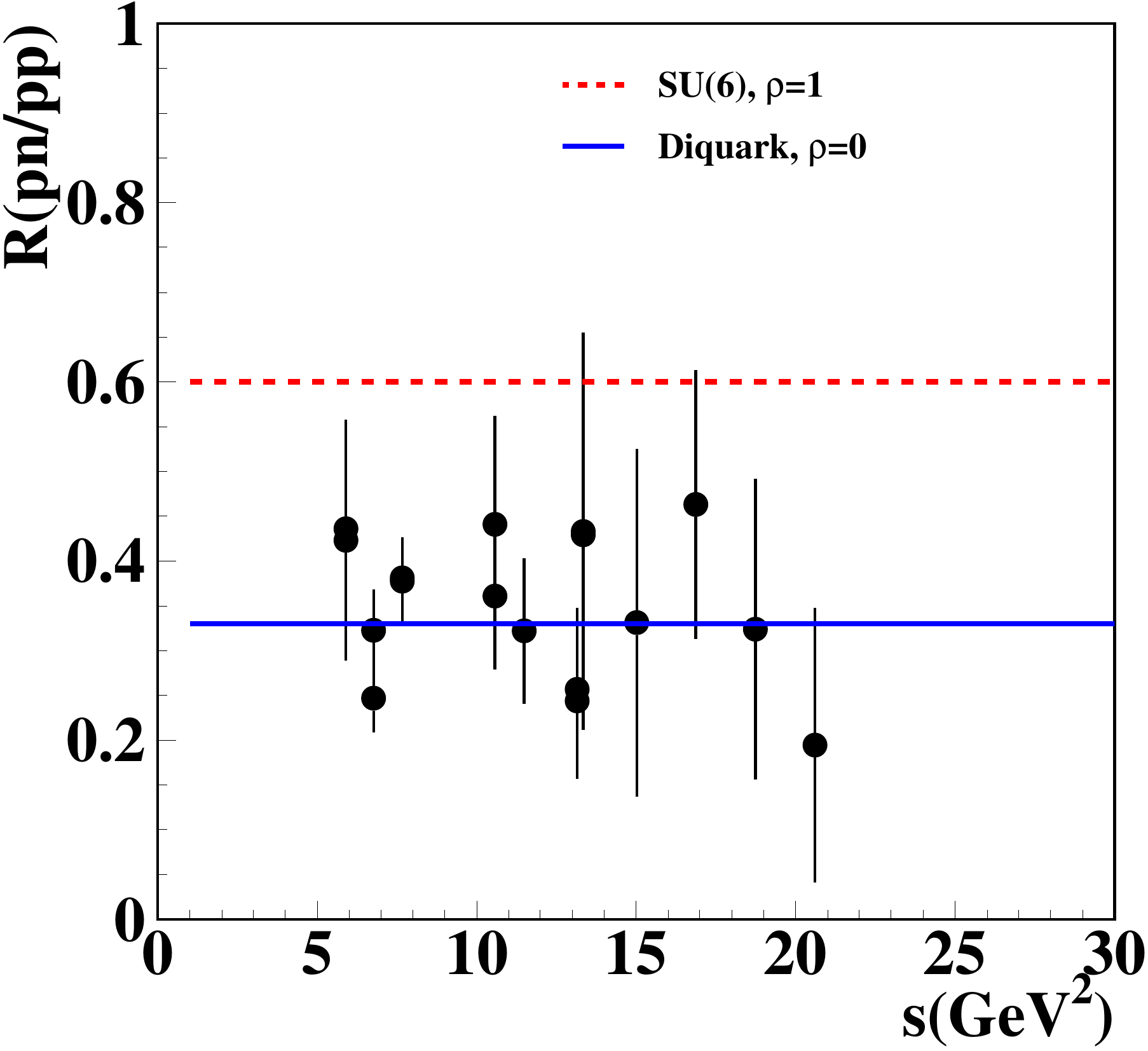}
\caption{(Color online) Ratio of the  $pn\rightarrow pn$ to $pp\rightarrow pp$   elastic differential cross sections 
as a function of $s$ at $\theta^{N}_{c.m.}=90^0$. }
\label{pn_to_pp_fig}
\end{figure}

As an illustration of the reliability of calculations of cross section ratios in the QI model, in Fig.\ref{pn_to_pp_fig} we compare the QI predictions for the ratios of $pn$ to $pp$ 
differential cross sections at $90^0$ c.m. scattering. Here, we compare predictions based on 
  SU(6)\cite{Farrar:1978by,Brodsky:1979nc} and diquark\cite{Granados:2009jh} symmetry assumptions for 
the valence quark wave function of the nucleons. As comparison shows one achieves  a rather reasonable 
agreement with the data without any additional normalization parameter. 
Based on this, we now estimate the ratio of the differential cross sections of $\gamma d\rightarrow \Delta\Delta$ 
to the $\gamma d \rightarrow pn$ cross sections. We use both SU(6) and diquark-symmetry quark wave functions  of the 
nucleon and  $\Delta$-isobars (see Appendix B) in the calculation of  the $pn\rightarrow \Delta\Delta$ amplitudes.

To calculate the photodisintegration amplitudes we go back to  Eqs.(\ref{asamp}) and (\ref{asampDD}) 
and evaluate the quark-charge factors using   SU(6) or diquark symmetries of the valence quark wave functions of baryons.  
For this  we separate the $t$ and $u$ channels in  the helicity amplitudes:
\begin{equation}
\phi_i(s,\theta_{c.m.}^N)= \phi^t_i(s,\theta_{c.m.}^N)+\phi^u_i(s,\theta_{c.m.}^N)
\end{equation}
and then treat the charge factors  for the given nucleon $N$ as:
\begin{equation}
\hat{Q}^{N}\phi_l=Q_i^{t,N}\phi^t_l+Q_i^{u,N}\phi^u.
\label{opa}
\end{equation}
This yields the following expression for the  photodisintegration amplitude of Eq.(\ref{ampl}) :
 \begin{eqnarray}
\langle\lambda_{1f},\lambda_{2f}\mid {\cal M} \mid \lambda_\gamma, \lambda_d\rangle 
& = & ie[\lambda_\gamma]\times \left\{\sum\limits_{\lambda_{2i}}  
{1\over \sqrt{2s'}}\left[Q^{tN_1}_i\phi^t_i+Q^{uN_1}_i\phi^u_i\right]_{\lambda_{2i}}
\int\Psi_{\textnormal{\scriptsize {d}}}^{\lambda_d}(p_1,\lambda_\gamma;p_2,\lambda_{2i})
{d^2p_\perp \over (2\pi)^2} \right. \nonumber \\ 
& + &\left.
\sum\limits_{\lambda_{1i}}{1\over \sqrt{2s'}}\left[Q^{tN_2}_i\phi^t_i+Q^{uN_2}_i\phi^u_i\right]_{\lambda_{1i}} \int 
\Psi_{\textnormal{\scriptsize {d}}}^{\lambda_d}(p_1,\lambda_{1i};p_2,\lambda_{\gamma})
{d^2p_\perp \over (2\pi)^2} \right\}.
\label{ampl5}
\end{eqnarray}

\subsection{$\gamma d\rightarrow pn$ scattering}
For the $\gamma d\rightarrow pn$ amplitude, the charge factors calculated for the helicity conserving amplitudes 
according to the QI framework yield for both SU(6) and diquark models (see Appendix B) 
\begin{eqnarray}
Q^{tN_1}_j & = & Q^{tN_2}_j={Q_{F,pn}\over2} \nonumber \\
Q^{uN_1}_j & = & -2Q^{uN_2}_j=2Q_{F,pn}
\label{QFpn}
\end{eqnarray}
 with $Q_{F,pn}={1\over3}$ and independent of j.  Using these relations in  Eq.(\ref{opa}),   from Eqs.(\ref{ampl5})  
and (\ref{asamp}) one obtains
\begin{equation}
|\bar{\cal M}|_{\gamma d\longrightarrow pn}^2={e^2\over 6\cdot 2 s^\prime }Q_{F,pn}^2\left\{S_{12}\phi_1^2+S_{34}
\left[\left({\phi^t_3+\phi^t_4\over2}+2\phi^u_4-\phi^u_3\right)^2+
\left({\phi^t_4+\phi^t_3\over2}+2\phi^u_3-\phi^u_4\right)^2\right]\right\},
\label{gdpnsa}
\end{equation}
where the  different predictions of SU(6) and diquark models follow from the different predictions for  the 
$pn\rightarrow pn$ helicity  conserving amplitudes given in Eq.(\ref{pnpn}). 

\subsection{$\gamma d\rightarrow \Delta^+\Delta^0$ scattering}
The calculation for  the $\gamma d\rightarrow \Delta^+\Delta^0$ amplitude yields the same quark-charge factors
as for the $\gamma d \rightarrow pn$ reactions in Eq.(\ref{QFpn}).  Using the  helicity amplitudes  of the 
$pn\rightarrow  \Delta^+\Delta^0$ scattering from Eq.(\ref{pnD+D0})  and  the expressions for 
the photodisintegration amplitudes from Eqs.(\ref{ampl5},\ref{asampDD})  one obtains
\begin{eqnarray}
|\bar{\cal M}|_{\gamma d\longrightarrow \Delta^+\Delta^-}^2&=&{1\over6}{e^2\over 2s^\prime} Q_{F,\Delta\Delta}^2\left\{S_{12}\left[|\phi_1|^2+|\phi_6|^2+|\phi_7|^2\right]\right.\nonumber\\&&
\left.+S_{34}\left[\left({\phi^t_3+\phi^t_4\over2}+2\phi^u_4-\phi^u_3\right)^2+\left({\phi^t_4+\phi^t_3\over2}+2\phi^u_3-\phi^u_4\right)^2\right.\right.\nonumber\\
&&\left.\left.+\left({\phi^t_8+\phi^t_9\over2}+2\phi^u_9-\phi^u_8\right)^2+\left({\phi^t_9+\phi^t_8\over2}+2\phi^u_8-\phi^u_9\right)^2\right]\right\},
\label{gdD+D0}
\end{eqnarray}
where the  different predictions of SU(6) and diquark models follow from the different predictions for the $pn\rightarrow \Delta^+\Delta^0$ helicity  
conserving amplitudes given in Eq.(\ref{pnD+D0}).

\subsection{$\gamma d\rightarrow \Delta^{++}\Delta^-$ scattering}

For the charge factors in  the  $\gamma d \rightarrow\Delta^{++}\Delta^-$ scattering  within the quark-interchange approximation  from 
Appendix B we obtain:
\begin{equation}
-Q^{tN_1}={Q^{tN_2}\over2}=Q_{F,\Delta\Delta}={1\over3}.
\end{equation}
Inserting these charge factors  in  Eqs.(\ref{ampl5},\ref{asampDD}) one obtains for the photodisintegration amplitude:
\begin{eqnarray}
|\bar{\cal M}|_{\gamma d\longrightarrow \Delta^{++}\Delta^-}^2= {1\over 6}
{e^2\over 2s^\prime}Q^2_{F,\Delta\Delta}\left\{S_{12}\left(|\phi_1|^2+
|\phi_6|^2+|\phi_7|^2\right)\right.\nonumber\\
+\left. S_{34}\left[\left(2\phi_3-\phi_4\right)^2+\left(2\phi_4-\phi_3\right)^2+5|\phi_8|^2\right]\right\}.
\label{gdD++D-}
\end{eqnarray}
where predictions for the  helicity conserving amplitudes of $pn\rightarrow \Delta^{++}\Delta^{-}$   are given in Eq.(\ref{pnD++D-}).

\subsection{Numerical Estimates}
Using Eqs.(\ref{gdpnsa}), (\ref{gdD+D0}) and (\ref{gdD++D-})  with the baryonic helicity amplitudes calculated in Appendix B 
we estimate the ratio $R(\theta_{c.m.})$   of the  $\gamma d\rightarrow\Delta\Delta$ 
to $\gamma d\rightarrow pn$  differential cross sections 
at given $s$ and  $\theta_{c.m.}$ angle.
 For simplicity we consider the kinematics  in which $s>>4m_\Delta^2$, which allows to approximate  both  Eqs.(\ref{theta_cmn}) and (\ref{kin2})  to,
\begin{eqnarray}
cos\theta^N_{c.m}\approx{1+cos\theta_{c.m.}\over2}.
\label{kin2a}
\end{eqnarray}
Before considering any specific model for angular distribution, one can make two general statements about the properties of the photodisintegration amplitude. First, that  from the absence of the $u$ channel  scattering in the $pn\rightarrow \Delta^{++}\Delta^{-}$ helicity amplitudes (see Eq.(\ref{pnD++D-})),
one observes that $R(\theta_{c.m.})$  can not be a uniform function of $\theta_{c.m}$.  Second, that  independent of the choice of SU(6) or diquark models,  
the $\gamma d \rightarrow \Delta^{++}\Delta^{-}$ cross section is always larger than the cross section of the $\gamma d\rightarrow \Delta^{+}\Delta^{-}$  
reaction.

We quantify the above observations by parameterizing  the angular function $f(\theta^N_{c.m.})$,  which enters in Eqs.(\ref{pnpn},\ref{pnD+D0},\ref{pnD++D-}),  in the
following form\cite{RS,Granados:2009jh}:
\begin{equation}
f(\theta) = {1\over sin(\theta)^2 (1-cos(\theta))^2}
\label{angf}
\end{equation}
known to describe reasonably well the elastic $pp$ and $pn$ scattering cross sections.

\begin{figure}[ht]
\centering\includegraphics[height=8cm,width=10cm]{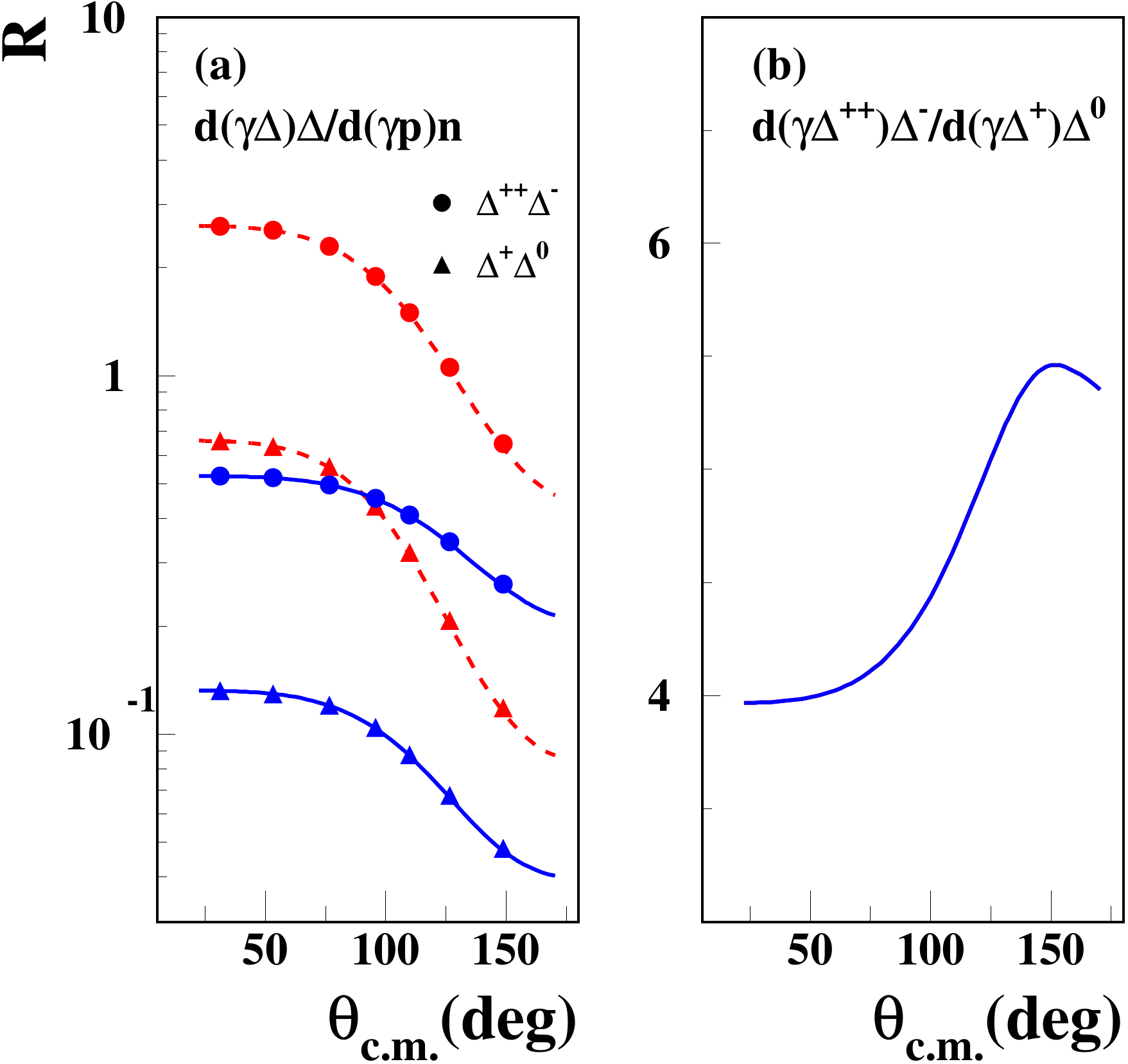}
\caption{(Color online) (a)Ratio of the  $\gamma d\rightarrow \Delta\Delta$ to $\gamma d\rightarrow pn$    differential cross sections and (b) ratio of the  $\gamma d\rightarrow \Delta^{++}\Delta^-$ to $\gamma d\rightarrow \Delta^+\Delta^0$    differential cross sections as a function of $\theta_{c.m.}$. }
\label{delte_delta_fig}
\end{figure}

Magnitudes of the ratio $R$ at $\theta_{c.m}=90^0$ are given in Table \ref{tab:t1}, while the angular dependencies (solid curves for diquark model and dashed curves for SU(6) model) are presented  in Fig.3(a). They clearly show strong angular anisotropy and  
the excess (by a factor of 4-5)  of the $\Delta^{++}\Delta^{-}$  breakup cross section relative to the cross section of the
 $\Delta^{+}\Delta^{0}$  breakup (Fig.3(b)). Our calculations show that the ratio of the $\gamma d\to \Delta\Delta$ to $\gamma d\to pn$ cross sections is very sensitive to the choice of SU(6) or diquark models of the wave functions. However, because of the absence of isosinglet two-quark state in the $\Delta$ wave functions, the $\rho$ parameter dependence that characterizes the choice of SU(6) or diquark models in the baryons wave functions is factorized and enters only in the normalization factor of the  $pn\to\Delta\Delta$ helicity  amplitudes. As a result, the ratio of the $\gamma d\to\Delta^{++}\Delta^-$ to $\gamma d\to \Delta^+\Delta^0$ cross sections (Fig.\ref{delte_delta_fig}b) is independent of the choice between SU(6) and diquark models for the baryons wave functions. 
 
\medskip
\medskip

Finally, it is worth discussing  how our calculations  compare with the predictions of models in which the production of  
two $\Delta$'s is a result of the breakup of the pre-existing $\Delta\Delta$ component of  the deuteron wave function.
In this case,  the final state interaction is dominated 
by soft scattering of two  $\Delta$'s in the final state  which  will induce   similar angular distributions for both 
$\Delta^{++}\Delta^{-}$ and $\Delta^{+}\Delta^{0}$ channels (see e.g.,\cite{FGMSS,gea}). As a result, we expect essentially the
same angular distribution for  both $\Delta^{++}\Delta^{-}$ and $\Delta^{+}\Delta^{0}$ production channels. Also, because of 
the deuteron being an isosinglet, the  probabilities of finding preexisting $\Delta^{++}\Delta^{-}$ and $\Delta^{+}\Delta^{0}$ are equal. 
For coherent hard  breakup of  the preexisting $\Delta$'s  we will obtain the  same cross section 
for both the  $\Delta^{++}\Delta^{-}$ and  the $\Delta^{+}\Delta^{0}$  channels.

\begin{table*}[htbp]
	\centering
		\begin{tabular}{c|c|c|} \cline{2-3} & \multicolumn{2}{|c|}{$R(90^o)$} \\ 
		\hline
		\multicolumn{1}{|c|}{$\gamma d\rightarrow BB$} &SU(6)&Diquark \\
		\hline
		\multicolumn{1}{|c|}{$\gamma d\rightarrow\Delta^+\Delta^0$}          &0.47 &0.11\\
		\hline
   \multicolumn{1}{|c|}{$\gamma d\rightarrow\Delta^{++}\Delta^-$}    &2.01 &0.47\\
		\hline 
		\end{tabular}
		\caption{Strength of $\Delta\Delta$ channels relative to $pn$ in deuteron photodisintegration at $\theta_{c.m}=90^o$.}
		\label{tab:t1}
\end{table*}

One interesting scenario for probing the preexisting $\Delta$'s in the deuteron  is 
using the decomposition of the  deuteron wave function, in the chiral symmetry restored limit, into the nucleonic 
and non-nucleonic components in the following form\cite{Harvey:1980rva,Brodsky:1983vf,Brodsky:1985gs}:
\begin{equation}
\Psi_{T=0,S=1} = ({1\over 9})^{1\over 2}\Psi_{NN} + ({4\over 45})^{1\over 2}\Psi_{\Delta\Delta} + ({4\over 5})^{1\over 2}\Psi_{CC},
\label{deutronCC}
\end{equation}
where $\Psi_{CC}$ represents the hidden color component of $T=0$ and $S=1$ six-quark configuration. 
Since $\Delta^{++}\Delta^{-}$ and $\Delta^{+}\Delta^{0}$ components enter with equal probability in 
the total isospin $T=0$ configuration, 
one expects close ($\approx 0.8$)  strengths for deuteron breakup to 
$\Delta^{++}\Delta^{-}$ or $\Delta^{+}\Delta^{0}$ channels as compared to the strength of the  deuteron breakup 
into the $pn$ pair. This result should be compared with the similar ratios presented in  Table \ref{tab:t1} from HRM
and with the HRM angular distributions in Fig.~\ref{delte_delta_fig}.

It is worth noting that HRM can be applied for calculation of the  large angle photo-production of  any given 
two baryonic resonances. In all cases the model will be sensitive to the valence quark wave function of the baryons 
as well as to the effective color charge factors entering in the scattering amplitude.   One such possibility is  the 
large center of mass angle photoproduction of the $NN^*$ pair  which will allow us to evaluate the 
role of the rescattering in reactions aimed at probing the $NN*$ component of the deuteron wave 
function.  Note that such a  process will not interfere with the  amplitude of $\Delta\Delta$  production at large 
center of mass angles, since the decay products of the produced resonances  occupy distinctly different phase 
spaces in the final state of the reaction.

\section{Summary}
We extended  the hard rescattering model of large c.m. angle photodisintegration of a two-nucleon system  to account for 
the production of two $\Delta$-isobars. The
HRM allows to express the cross section of  $\gamma d\rightarrow pn$ and $\gamma d\rightarrow \Delta\Delta$ reactions through the 
large c.m. angle differential cross section of $pn\rightarrow pn$ and $pn\rightarrow \Delta\Delta$   scattering amplitudes.

Because of  lack of  experimental information on $pn\rightarrow \Delta\Delta$   scattering, we further applied 
the quark-interchange model to calculate the  strength of the  $\gamma d \rightarrow \Delta\Delta$  cross section relative to the 
cross section of $\gamma d\rightarrow pn$ breakup reaction.   We predicted a significantly larger strength for 
 the $\Delta^{++}\Delta^{-}$  channel of breakup as compared to the $\Delta^{+}\Delta^{0}$ channel which is related to the 
relative strength of the  $pn\rightarrow \Delta^{++}\Delta^{-}$ and  $pn\rightarrow \Delta^{+}\Delta^{0}$ scatterings.
Because of the different angular dependences of these hadronic amplitudes, we also predicted a significant difference between the
angular dependences of  photoproduction cross sections in  $\Delta^{++}\Delta^{-}$ and $\Delta^{+}\Delta^{0}$ channels.

These results can be compared with the prediction of the models in which two  $\Delta$'s are produced due to the coherent 
breakup of the $\Delta\Delta$ component of the deuteron wave function.  In this case one expects essentially similar angular 
distributions and  strengths for  the $\Delta^{++}\Delta^{-}$ and $\Delta^{+}\Delta^{0}$ breakup channels.

\acknowledgments
We are grateful to  Drs. Stanley Brodsky, Haiyan Gao, Patrizia Rossi and Mark Strikman for 
many useful discussions and comments.  This work is supported by U.S. Department of Energy Grant 
under Contract DE-FG02-01ER41172 and by
Florida International University's Doctoral Evidence
Acquisition program.

\appendix

 \section{Baryon-Baryon Scattering Helicity Amplitudes}
\label{bbsa}
We are using helicity states to label the entries of the photodisintegration and the baryon-baryon scattering matrices. 
The  number of independent helicity  amplitudes for a given $ab\rightarrow cd$ processes can be expressed through the total 
spin of the scattering particles as follows\cite{perl:1974,Perl:1969pg}: 
\begin{eqnarray}
N={1\over 2}\cdot (2s_a+1)(2s_b+1)(2s_c+1)(2s_d+1)
\label{nampl}
\end{eqnarray}
where $s_i$ is the total spin of particle i and  for the  photon  we replace $(s_i+1)$ by 2.   The factor ${1\over 2}$  
follows from the  constraint due to the  parity conservation.
For elastic scattering, there is a further reduction in $N$ due to time reversal invariance, and if the scattering particles are identical, or lie in the 
same isospin multiplet, the number of independent helicity amplitudes is reduced  further\cite{perl:1974,Perl:1969pg}. 
For the $pn$ elastic scattering case, out of the possible 16 helicity amplitudes only five  are independent\cite{Perl:1969pg} for which we use the 
following notations:
\begin{eqnarray}
\left\langle +{1\over2},+{1\over2} \left|T\right|+{1\over2},+{1\over2}\right\rangle &=& \phi_1\nonumber\\
\left\langle +{1\over2},-{1\over2} \left|T\right|+{1\over2},-{1\over2}\right\rangle &=& \phi_3\nonumber\\
\left\langle -{1\over2},+{1\over2} \left|T\right|+{1\over2},-{1\over2}\right\rangle &=& \phi_4\nonumber\\
\left\langle -{1\over2},-{1\over2} \left|T\right|+{1\over2},+{1\over2}\right\rangle &=& \phi_2\nonumber\\
\left\langle -{1\over2},+{1\over2} \left|T\right|+{1\over2},+{1\over2}\right\rangle &=& \phi_5,\nonumber\\
\label{pnham}
\end{eqnarray}

For  the $pn\rightarrow\Delta\Delta$ scattering amplitude, we have from Eq.(\ref{nampl}), $N$=(2)(2)(4)(4)/2=32 
independent helicity amplitudes.  We use the following notations for  the helicity conserving independent 
amplitudes of  $pn\rightarrow\Delta\Delta$ scattering:
 \begin{eqnarray}
\left\langle +{1\over2},+{1\over2} \left|T\right|+{1\over2},+{1\over2}\right\rangle &=& \phi_1\nonumber\\
\left\langle +{1\over2},-{1\over2} \left|T\right|+{1\over2},-{1\over2}\right\rangle &=& \phi_3\nonumber\\
\left\langle -{1\over2},+{1\over2} \left|T\right|+{1\over2},-{1\over2}\right\rangle &=& \phi_4\nonumber\\
\left\langle +{3\over2},-{1\over2} \left|T\right|+{1\over2},+{1\over2}\right\rangle &=& \phi_6\nonumber\\
\left\langle -{1\over2},+{3\over2} \left|T\right|+{1\over2},+{1\over2}\right\rangle &=& \phi_7\nonumber\\
\left\langle +{3\over2},-{3\over2} \left|T\right|+{1\over2},-{1\over2}\right\rangle &=& \phi_8\nonumber\\
\left\langle -{3\over2},+{3\over2} \left|T\right|+{1\over2},-{1\over2}\right\rangle &=& \phi_9,\nonumber\\
\label{hconamp}
\end{eqnarray}
which are consistent with the definitions in Eq. (\ref{pnham}).

\section{Helicity Amplitudes of Photodisintegration in  the Quark-Interchange Model}

\subsubsection{Quark Interchange model}
Following the approach presented for example in Refs.\cite{Sivers:1975dg,Farrar:1978by,Brodsky:1979nc, Granados:2009jh}, 
the scattering amplitude for a process $ab\rightarrow cd$, in which $a,b,c$ and $d$ are baryons, 
is obtained from,
\begin{eqnarray}
& & \langle cd\mid T\mid ab\rangle = 
\sum\limits_{\alpha,\beta,\gamma} 
\langle  \psi^\dagger_c\mid\alpha_2^\prime,\beta_1^\prime,\gamma_1^\prime\rangle
\langle  \psi^\dagger_d\mid\alpha_1^\prime,\beta_2^\prime,\gamma_2^\prime\rangle 
\nonumber \\
& & \ \ \ \  \times
\langle \alpha_2^\prime,\beta_2^\prime,\gamma_2^\prime,\alpha_1^\prime\beta_1^\prime
\gamma_1^\prime\mid H\mid
\alpha_1,\beta_1,\gamma_1,\alpha_2\beta_2\gamma_2\rangle\cdot 
\langle\alpha_1,\beta_1,\gamma_1\mid\psi_a\rangle
\langle\alpha_2,\beta_2,\gamma_2\mid\psi_b\rangle,
\label{qimampl}
\end{eqnarray}
where ($\alpha_i,\alpha_i^\prime$), ($\beta_i,\beta_i^\prime$) and 
($\gamma_i\gamma_i^\prime$) describe the spin-flavor  quark states 
before and after the hard scattering, $H$,
and
\begin{equation}
C^{j}_{\alpha,\beta,\gamma} = \langle\alpha,\beta,\gamma\mid\psi_j\rangle
\label{Cs}
\end{equation}
describes the probability amplitude of finding an $\alpha,\beta,\gamma$ helicity-flavor 
combination of three valence quarks in the baryon $j$.
These coefficients are obtained from the expansion of the 
baryon's spin-isospin wave function in three-quark valence states as follows:
\begin{eqnarray}
\psi^{i^3_{N},h_N} & = & {N\over \sqrt{2}}\left\{
\sigma (\chi_{0,0}^{(23)}\chi_{{1\over2},h_N}^{(1)})\cdot
(\tau_{0,0}^{(23)}\tau_{{1\over 2},i_N^{3}}^{(1)}) 
\right.  +   \nonumber \\
& & 
\rho \sum\limits_{i_{23}^3=-1}^{1} \ \ \sum\limits_{h_{23}^3=-1}^{1}
\langle 1,h_{23}; {1\over 2},h_{N}-h_{23}\mid {1\over 2},h_N\rangle
\langle 1,i^3_{23}; {1\over 2},i^3_{N}-i^3_{23}\mid {1\over 2},i^3_N\rangle \nonumber \\
& &\left. \times (\chi_{1,h_{23}}^{(23)}\chi_{{1\over2},h_N-h_{23}}^{(1)})\cdot
(\tau_{1,i^3_{23}}^{(23)}\tau_{{1\over 2},i_N^{3}-i^3_{23}}^{(1)})\right\}.
\label{wf}
\end{eqnarray}
The indexes 1 and 23  label the quark and the diquark states. The first term corresponds to quarks 2 and 3 being in a helicity zero isosinglet state, while the second term corresponds to quarks 2 and 3 in  helicity 1-isotriplet states. 
Where $\chi$ and $\tau$ represent helicity and isospin states with helicity $h$ and isospin projection $i^3$ respectively.    For the wave functions of $\Delta$-isobars $\sigma=0$ and $\rho=1$, while for nucleon 
wave functions $\sigma=1$ and the parameter $\rho$  characterizes the average  strength of the isotriplet diquark radial state relative 
to that of the isosinglet state.  Two extreme values of $\rho=1$ and $\rho=0$ correspond to the 
realization of the SU(6) and good diquark symmetries in the wave function. 
 
Using Eq.(\ref{wf}) in Eq.(\ref{qimampl}) for the hadronic scattering amplitude  one obtains:
 
\begin{equation}
\langle cd |T^{QIM}|ab\rangle = A_{\alpha_1',\alpha_2',\alpha_1\alpha_2}(\theta^N_{c.m.})
M^{ac}_{\alpha_1,\alpha_1'}M^{bd}_{\alpha_2,\alpha_2'} + A_{\alpha_1',\alpha_2',\alpha_1\alpha_2}(\pi-\theta^N_{c.m.}) 
M^{ad}_{\alpha_1,\alpha_1'}M^{bc}_{\alpha_2,\alpha_2'},
\label{elamp}
\end{equation} 
where
\begin{equation}
M^{ij}_{\alpha,\alpha'} = C^{i}_{\alpha,\beta,\gamma}C^{j}_{\alpha',\beta,\gamma} +
C^{i}_{\beta,\alpha,\gamma}C^{j}_{\beta,\alpha',\gamma} + 
C^{i}_{\beta,\gamma,\alpha}C^{j}_{\beta,\gamma,\alpha'},  
\label{qimamp}
\end{equation}
which accounts for all possible interchanges of $\alpha$ and $\alpha'$ quarks leaving $\beta$ and $\gamma$ quarks unchanged. In the QI model
 the interchanging quarks conserve their corresponding helicities and flavors, this is accounted for in the matrix elements of $A$ in Eq.(\ref{elamp}.),
\begin{equation}
 A_{\alpha_1',\alpha_2',\alpha_1\alpha_2}(s,\theta^N_{c.m.}) \propto 
\delta_{\alpha_1',\alpha_2}\delta_{\alpha_2',\alpha_1} 
{f(\theta^N_{c.m.})\over s^2}
\label{qimfw}
\end{equation}

Eq.(\ref{elamp}) has two terms, first (referred as a $t$ term)  in which four quarks  scatter at  angle $\theta^N_{c.m.}$ and two (interchanging) quarks scatter 
at $\pi-\theta^N_{c.m.}$  and the second (referred as a $u$ term) in which two interchanging quarks scatter at $\theta^N_{c.m.}$, while  four spectator quarks scatter 
at $\pi-\theta^N_{c.m.}$.

\subsection{Helicity Amplitudes in the Quark Interchange Model}
Through the above procedure using Eq.(\ref{elamp}) for the helicity amplitudes of $pn$ scattering one obtains:
\begin{eqnarray}
\phi_1(\theta^N_{c.m.})&=&(2-y)f(\theta^N_{c.m.})+(1+2y)f(\pi-\theta^N_{c.m.})\\\nonumber
\phi_2(\theta^N_{c.m.})&=&0\\\nonumber
\phi_3(\theta^N_{c.m.})&=&(2+y)f(\theta^N_{c.m.})+(1+4y)f(\pi-\theta^N_{c.m.})\\\nonumber
\phi_4(\theta^N_{c.m.})&=&2yf(\theta^N_{c.m.})+2yf(\pi-\theta^N_{c.m.})\\\nonumber
\phi_5(\theta^N_{c.m.})&=&0,
\label{pnpn}
\end{eqnarray}
were,
\begin{eqnarray}
y&=&{2\over3}{\rho\over{1+\rho^2}}\left(1+{2\over3}{\rho\over{1+\rho^2}}\right).
\end{eqnarray}

For $pn\rightarrow\Delta^+\Delta^0$ scattering amplitudes we obtain:
\begin{eqnarray}
\phi_1&=&{2\over9}N_{\Delta\Delta}(2f(\theta^N_{c.m.})-f(\pi-\theta^N_{c.m.}))\nonumber\\
\phi_3&=&{1\over9}N_{\Delta\Delta}(4f(\theta^N_{c.m.})+f(\pi-\theta^N_{c.m.}))\nonumber\\
\phi_4&=&{2\over9}N_{\Delta\Delta}(f(\theta^N_{c.m.}))+f(\pi-\theta^N_{c.m.})\nonumber\\
\phi_6&=&{N_{\Delta\Delta}\over3\sqrt{3}}(2f(\theta^N_{c.m.})-f(\pi-\theta^N_{c.m.}))\nonumber\\
\phi_7&=&{N_{\Delta\Delta}\over3\sqrt{3}}(2f(\theta^N_{c.m.})-f(\pi-\theta^N_{c.m.}))\nonumber\\
\phi_8&=&{2\over9}N_{\Delta\Delta}f(\theta^N_{c.m.})\nonumber\\
\phi_9&=&{1\over3}N_{\Delta\Delta}f(\pi-\theta^N_{c.m.}),\nonumber\\
\label{pnD+D0}
\end{eqnarray}
and similarly for the  amplitudes of the $pn\rightarrow\Delta^{++}\Delta^-$ scattering, QI model gives:
\begin{eqnarray}
\phi_1&=&-{2\over3}N_{\Delta\Delta}f(\theta^N_{c.m.})\nonumber\\
\phi_3&=&-{2\over3}N_{\Delta\Delta}f(\theta^N_{c.m.})\nonumber\\
\phi_4&=&-{1\over3}N_{\Delta\Delta}f(\theta^N_{c.m.})\nonumber\\
\phi_6&=&{-N_{\Delta\Delta}\over\sqrt{3}}f(\theta^N_{c.m.})\nonumber\\
\phi_7&=&{-N_{\Delta\Delta}\over\sqrt{3}}f(\theta^N_{c.m.})\nonumber\\
\phi_8&=&-N_{\Delta\Delta}f(\theta^N_{c.m.})\nonumber\\
\phi_9&=&0,\nonumber\\
\label{pnD++D-}
\end{eqnarray}
For both sets of equations in  (\ref{pnD+D0}) and (\ref{pnD++D-}), we have
\begin{equation}
N_{\Delta\Delta}={(1+\rho)^2\over1+\rho^2},
\label{N_DD}
\end{equation}
which shows that the  strength of the two $\Delta\Delta$ channels relative to each other is independent of the value of $\rho$. This is not the case for their strengths  relative to the $pn$ channel; from Eqs. (\ref{pnpn}) we see that the $\rho$ dependence of the helicity amplitudes in $pn\rightarrow pn$ cannot be factorized.

\subsection{Quark-Charge Factors}

 In the hard rescattering model, photodisintegration amplitudes are expressed in terms of hadronic scattering amplitudes weighted by the charges of 
struck quarks, Eq.(\ref{QT}).   We further split the amplitude of Eq.(\ref{QT}) into $t$ and $u$ channel scatterings:  
\begin{eqnarray}
\sum_{i}Q_i^{N_k}\langle \lambda_{2f};\lambda_{1f}\mid T_{(pn\rightarrow B_1B_2),i}(s,\tilde{t})\mid \lambda_\gamma;\lambda_{2i}\rangle=\left[Q^{tN_k}_j\phi^t_j+Q^{uN_k}_j\phi^u_j\right],
\label{emqim}
\end{eqnarray}
where $Q^{t/uN}_i$ is the charge of the quark,  struck  by the incoming photon from the nucleon $N$ with further $\theta^N_{c.m.}$ or $\pi-\theta^N_{c.m.}$  scattering.  
The helicity amplitudes are also split into $t$ and $u$ parts 
 \begin{eqnarray}
 \phi_i(\theta^N_{c.m.})&=& \phi^t_i(\theta^N_{c.m.})+\phi^u_i(\theta^N_{c.m.})\nonumber\\
 &=& c_tf(\theta^N_{c.m.})+c_uf(\pi-\theta^N_{c.m.}),
 \label{tuqim}
 \end{eqnarray}
with  $\phi^t$ and $\phi^u$ corresponding to the $\theta^N_{c.m.}$ or $\pi-\theta^N_{c.m.}$  scattering terms in Eq.(\ref{elamp}).

Using the above definitions and Eqs.(\ref{elamp},\ref{qimamp},\ref{pnpn},\ref{pnD+D0},\ref{pnD++D-}) 
the charge factors $Q^t$ and $Q^u$ are calculated using 
the following relations:
\begin{eqnarray}
Q^{tN_k}_j&=&{Q(\alpha_k)A_{\alpha_1',\alpha_2',\alpha_1\alpha_2} 
M^{ac}_{\alpha_1,\alpha_1'}M^{bd}_{\alpha_2,\alpha_2'}\over \phi^t_j}\nonumber\\
Q^{uN_k}_j&=&{Q(\alpha_k)A_{\alpha_1',\alpha_2',\alpha_1\alpha_2} 
M^{ad}_{\alpha_1,\alpha_1'}M^{bc}_{\alpha_2,\alpha_2'}\over \phi^u_j},
\label{chfctr}
\end{eqnarray}
where summation is understood for repeated $\alpha$ indices, $Q(\alpha)$ is the charge in e units of a quark $\alpha$ and the index $ j$ labels the 
process $ab\rightarrow cd$.

\end{document}